# Designing magnetocaloric materials for hydrogen liquefaction with light rare-earth Laves phases

Wei Liu[1,*], Tino Gottschall[2], Franziska Scheibel[1], Eduard Bykov[2,3], Nuno Fortunato[1], Alex Aubert[1], Hongbin Zhang[1], Konstantin Skokov[1], and Oliver Gutfleisch[1]

[1]Institute of Materials Science, Technical University of Darmstadt, 64287 Darmstadt, Germany
[2]Dresden High Magnetic Field Laboratory (HLD-EMFL) and Würzburg-Dresden Cluster of Excellence ct.qmat, Helmholtz-Zentrum Dresden-Rossendorf, 01328 Dresden, Germany
[3]Institute of Solid State and Materials Physics, Technische Universität Dresden, 01062 Dresden, Germany

[*]Author to whom correspondence should be addressed: wei.liu@tu-darmstadt.de Alarich-Weiss-str. 16, 64287 Darmstadt

**Abstract**

Magnetocaloric hydrogen liquefaction could be a "game-changer" for liquid hydrogen industry. Although heavy rare-earth based magnetocaloric materials show strong magnetocaloric effects in the temperature range required by hydrogen liquefaction (77 ~ 20 K), the high resource criticality of the heavy rare-earth elements is a major obstacle for upscaling this emerging liquefaction technology. In contrast, the higher abundances of the light rare-earth elements make their alloys highly appealing for magnetocaloric hydrogen liquefaction. Via a mean-field approach, it is demonstrated that tuning the Curie temperature ($T_C$) of an idealized light rare-earth based magnetocaloric material towards lower cryogenic temperatures leads to larger maximum magnetic and adiabatic temperature changes ($\Delta S_T$ and $\Delta T_{ad}$). Especially in the vicinity of the condensation point of hydrogen (20 K), $\Delta S_T$ and $\Delta T_{ad}$ of the optimized light rare-earth based material are predicted to show significantly large values. Following the mean-field approach and taking the chemical and physical similarities of the light rare-earth elements into consideration, a method of designing light rare-earth intermetallic compounds for hydrogen liquefaction is used: tunning $T_C$ of a rare-earth alloy to approach 20 K by mixing light rare-earth elements with different *de Gennes* factors. By mixing Nd and Pr in Laves phase $(Nd,Pr)Al_2$, and Pr and Ce in Laves phase $(Pr,Ce)Al_2$, a fully light rare-earth intermetallic series with large magnetocaloric effects covering the temperature range required by hydrogen liquefaction is developed, demonstrating a competitive maximum effect compared to the heavy rare-earth compound $DyAl_2$.

## 1. Introduction

Discovered in 1917 by Weiss and Picard, magnetocaloric effect is a cooling or warming effect of a magnetic material being exposed to a magnetic field [1]. Soon after its discovery, magnetic cooling has been successfully applied to attaining extremely low temperature [2, 3]. In 1949, the Nobel prize in chemistry was awarded to Giauque, who developed a magnetic refrigeration device to approach absolute zero [4].

Nowadays, global climate change caused by greenhouse-gas emissions threatens human civilization with grave consequences. In order to achieve climate-neutrality, new energy concepts emphasizing technologies that improve energy efficiency, or replace fossil fuels are required [5]. Ideally releasing no pollutants and greenhouse gases, green hydrogen is finally set to be the fuel for the future [6–8]. Hydrogen liquefaction is important for efficient storage and transportation of hydrogen energy [7, 9, 10]. However, liquid hydrogen is expensive due to the low efficiency of the conventional liquefaction technologies based on Joule-Thomson expansion [11, 12]. Recently, there is a growing interest in magnetocaloric liquefaction for hydrogen and other industrial gases [13–25]. The emerging magnetocaloric liquefaction technology is in principle more efficient [5, 26–32], making the promise of hydrogen fuel being affordable for the society to reach climate-neutrality.

If pre-cooled by liquid nitrogen, the temperature range required by magnetocaloric hydrogen liquefaction is from 77 (condensation point of nitrogen) to 20 K (condensation point of hydrogen). For the success of a practical application of magnetocaloric hydrogen liquefaction on an industrial scale, affordable magnetocaloric materials with large isothermal magnetic entropy and adiabatic temperature changes ($\Delta S_T$ and $\Delta T_{ad}$) in the target temperature regime under affordable magnetic fields are needed [33–38]. In this work, we focus on the criticality of raw elements, the two physical quantities $\Delta S_T$ and $\Delta T_{ad}$ in temperature range of 77 ~ 20 K under magnetic field changes such as 2 T, which can be realized by Nd-Fe-B permanent magnets [5], or 5 T, which can be generated by commercial superconducting magnets [32].

Rare-earth based magnetocaloric materials are one big family of the magnetocaloric materials for hydrogen liquefaction [37, 39, 40]. Lanthanide rare-earth elements can be divided into two major groups: light rare-earth elements (La, Ce, Pr, Nd, Sm) and heavy rare-earth elements (Gd, Tb, Dy, Ho, Er, Tm, Yb, Lu). Putting aside the criticality of the raw materials for a moment, heavy rare-earth based magnetocaloric materials would be strong contenders for magnetocaloric hydrogen liquefaction in the context of performance. Materials such as HoB$_2$ [15, 41, 42], ErCo$_2$ [43, 44], DyAl$_2$ [45], and ErAl$_2$ [45, 46] show large $\Delta S_T$ and $\Delta T_{ad}$ due to the large magnetic moments of the heavy rare-earth ions, and they are often proposed to be used in an active magnetic regenerator for hydrogen liquefaction. Because of the excellent magnetocaloric properties, heavy rare-earth based materials are intensively studied.

In contrast, light rare-earth based magnetocaloric materials for hydrogen liquefaction are less investigated since their magnetocaloric effects are usually weaker due to their smaller magnetic moments [45]. **Figure 1** (a) shows the theoretical effective magnetic moment $\mu_{eff}$ and the *de Gennes* factor ($G$) of the rare-earth ions [47]. The rare-earth ions are divided into three categories, namely the light rare-earth ions, Eu$^{3+}$, and the heavy rare-earth ions. Pr$^{3+}$ has the highest theoretical $\mu_{eff}$ of 3.58 $\mu_B$ among the light rare-earth ions, whereas the heavy rare-earth ions from Gd$^{3+}$ to Tm$^{3+}$ have a theoretical $\mu_{eff}$ larger than 7.5 $\mu_B$.

However, the criticality of the raw materials cannot be ignored for a viable industrial-scale application. The consumption of H$_2$ in EU is predicted to reach 2250 TWh/year by 2050 according to Hydrogen Roadmap Europe [48]. If about one-third of H$_2$ needs to be transported and stored in its liquid state, one would require about 13,000 small-scale liquefaction plants with a production capacity of 5 tons per day. Providing a potential hydrogen liquefier using

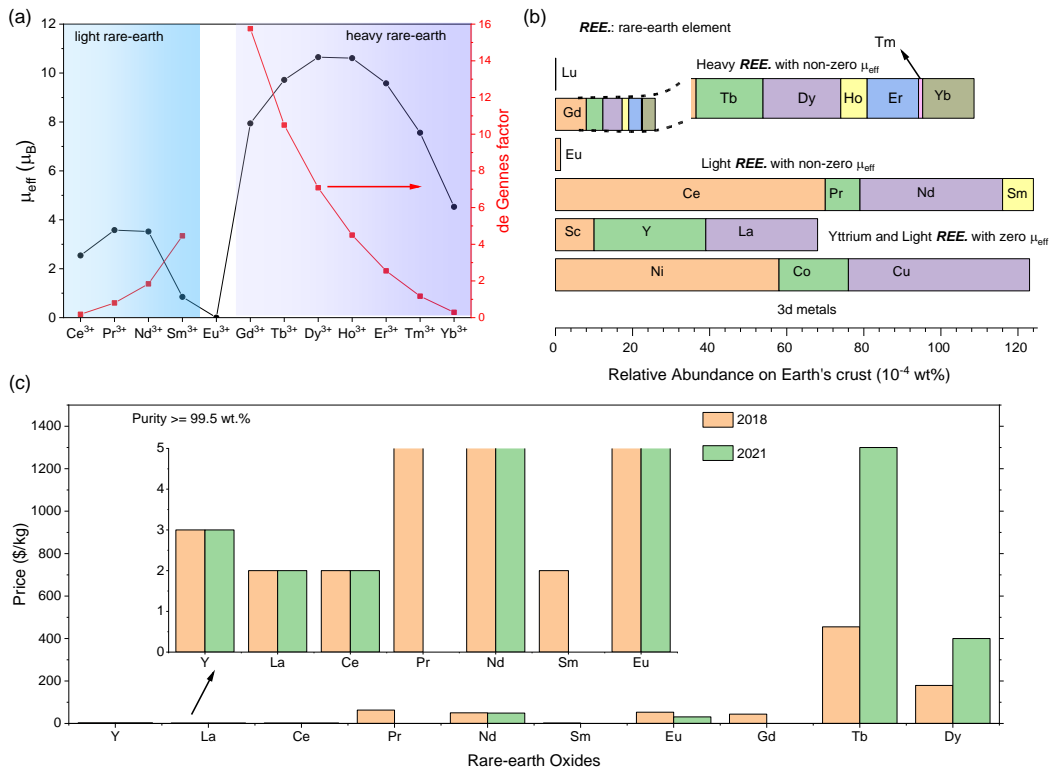

**Figure 1** (a) Effective magnetic moments and *de Gennes* factors of the rare-earth ions [47]. (b) relative abundance of Ni, Co, Cu, and the rare-earth elements which are categorized into non-heavy rare-earth elements with zero magnetic moments (Sc, Y, La), light rare-earth elements with non-zero magnetic moments (Ce, Pr, Nd, Sm), Eu, and heavy rare-earth elements with non-zero magnetic moments (Gd, Tb, Dy, Ho, Er, Tm, Yb), and zero magnetic moments (Lu). Data are taken from Ref.[53]. (c) Prices of the rare-earth oxides (Y, La, Ce, Pr, Nd, Sm, Eu, Gd, Tb, Dy) in year 2018 and 2021. Data are taken from Ref. [54, 55].

magnetocaloric material $HoAl_2$ as the refrigerant for the final cooling state operating at 20 K at a frequency of 1 Hz in fields of 7 T, about 1 ton of holmium would be needed [49]. This means a total of 13,000 tons for the EU alone, and the total holmium production is a measly 10 tons per year [50].

Heavy rare-earth elements belong to the highly critical elements [51, 52]. One contribution to the high criticalities is their poor abundances on the earth's crust. **Figure 1** (b) shows the relative abundances of Ni, Co, Cu, and the rare-earth elements on the earth's crust [53]. Heavy rare-earth elements such as Tb, Ho, Tm, and Lu are not abundant, whereas light rare-earth metal Ce is even more abundant than Cu, and light rare-earth elements La and Nd, and Y are more abundant than Co. The total abundances of the heavy rare-earth elements combined are not as rich as that of Nd alone. **Figure 1** (c) plots the prices (always volatile) of the rare-earth oxides in year 2018 and 2021 for Y, La, Ce, Pr, Nd, Sm, Eu, Gd, Tb, and Dy [54, 55]. The prices of the heavy rare-earth oxides, namely for Tb, Dy, are over 400 \$/kg in 2021, whereas the prices of the Pr and Nd oxides are around 50 \$/kg, and the prices of Y, La, Ce, Sm oxides are just several dollars per kg. Eu and Gd oxides are also relatively cheap in comparison with Tb and Dy oxides. This is because they are not as largely used in industry as Tb and Dy.

It needs to be emphasized that criticality is much more than just simple geological abundances. Factors such as mining, beneficiation, hazardous by-products, separation (and their social and ecological consequences along this value chain), geopolitics, trade restrictions and monopolistic supply in terms of demand vs. supply, need to be understood and quantified in terms of LCA (Life-Cycle-Analysis) and LCC (Life-Cycle-Costing) [56]. Nevertheless, the discussion above points out that the high criticality of the heavy rare-earth elements questions the feasibility of using heavy rare-earth based magnetocaloric materials for hydrogen liquefaction in a viable industrial scale [57, 58]. Taking the dominant abundance of the light rare-earth elements over the heavy rare-earth elements into consideration, it is more feasible to use light rare-earth based magnetocaloric materials for hydrogen liquefaction. In this work, we aim at developing a light rare-earth based material system with sufficient magnetocaloric effects covering a full temperature range (77 ~ 20 K) required by magnetocaloric hydrogen liquefaction.

It is worth mentioning that much work has been done on tuning the transition temperature of heavy rare-earth magnetocaloric materials. Mixing two different heavy rare-earth elements on the rare-earth sites is a widely used method. Examples are several works on heavy rare-earth Laves phases $RNi_2$, $RAl_2$, and $RCo_2$ (R: rare-earth element) systems: the $Dy_{1-x}Er_xNi_2$ (x = 0.25, 0.5, 0.75) [59], the $Tb_{1-x}Er_xNi_2$ (x = 0.75, 0.5, 0.25)[60], the $Tb_{1-x}Ho_xNi_2$ (x = 0.25, 0.5, 0.75)[61], the $Ho_{1-x}Er_xNi_2$ (x =0.25, 0.5, 0.75 )[62], the $Tm_xDy_{1-x}Al_2$ (0≤x≤1)[13], the $(Er_xR_{1-x})Co_2$ (R=Ho, Dy; 0≤x≤1) [63] and the $Er_xDy_{1-x}Al_2$ (x = 0.45, 0.67, 0.9) [64]. Following these studies, we apply this method to the light rare-earth Laves phases: mixing different light rare-earth elements on the rare-earth sites to tune the transition temperature within 20 ~ 77 K for magnetocaloric hydrogen liquefaction.

## 2. Mean-field Approach

Inspired by the sharply increasing trends of $\Delta S_T$ and $\Delta T_{ad}$ with decreasing $T_C$ in the vicinity of the condensation point of hydrogen (20 K) for heavy rare-earth based magnetocaloric materials demonstrated by a mean-field approach from our previous work (Ref. [45] ), we focus in this work on the light rare-earth based magnetocaloric materials. We continue to develop the mean-field approach used in the aforementioned work, where the detailed description can be found [45]. It is worth mentioning that using the mean-field model to describe the magnetocaloric properties is a well-known method, as used in Ref. [65–69]. In some studies, also effects related to the crystalline electric field are taken into consideration [70–72]. In this work, we aimed at providing a simple way to understand the sharply increasing feature of magnetocaloric effect of the light rare-earth magnetocaloric materials with a Curie temperature in the vicinity of hydrogen condensation point.

The total entropy change $S$ of a magnetocaloric material is contributed by three items, namely the magnetic entropy $S_m$, the lattice entropy $S_l$, and the electronic entropy $S_e$:
$$S = S_m + S_l + S_e, \quad (1)$$

The equation to calculate the magnetic entropy is given as [73]:
$$S_m = N_M k_B \left[ \ln\left(\frac{\sinh\frac{2J+1}{2J}x}{\sinh\frac{1}{2J}x}\right) - x B_J(x) \right], \quad (2)$$

with $x = \frac{\mu\mu_0 H + \frac{3J}{J+1}k_B T_C B_J(x)}{k_B T}$. $H$ is the magnetic field, $J$ the total angular momentum, $N_M$ the number of "magnetic atoms", $k_B$ the Boltzmann constant, $\mu$ the atomic magnetic moment, $T_C$ the Curie temperature, $\mu_0$ the vacuum permeability, and $B_J(x)$ the Brillouin function. A more detailed description on Equation (2) can be found in Ref. [73]. The lattice entropy is given as [73]:
$$S_l = -3Nk_B\left[1 - \exp\left(-\frac{T_D}{T}\right)\right] + 12Nk_B\left(\frac{T}{T_D}\right)^3 \int_0^{\frac{T_D}{T}} \frac{x^3}{\exp(x)-1} dx. \quad (3)$$

where $T_D$ is the Debye temperature, $N$ the total number of atoms, and $x$ can be regarded as a variable in the range of 0 ~ $T_D/T$. The electronic entropy is given by [73]

$$S_e = \int_0^T \frac{C_e}{T} dT = \gamma T,  \quad (4)$$

where $\gamma$ is the Sommerfeld coefficient and $C_e$ is the electronic heat capacity ($C_e = \gamma T$). In the present work, $S_e$ is neglected out of simplification, since mostly $C_e$ is dominant only at sufficiently low temperatures [74].

Constructing the total entropy $S(T,H)$ curves by summing $S_m$ and $S_l$, the isothermal magnetic and adiabatic temperature changes ($\Delta S_T$ and $\Delta T_{ad}$) are given by [73]

$$\Delta S_T(T,H) = S(T,H) - S(T,0),$$
$$\Delta T_{ad}(T,H) = T(S,H) - T(S,0), \quad (5)$$

where $T(S,H)$ is the inverse function of $S(T,H)$. More details on Equation (1)~(5) can be found in Ref. [73].

We assume an idealized Nd- alloy family with the $T_C$ of its alloys varying from 300 K to 10 K and the other parameters, namely $J$, $\mu_{eff}$, and $T_D$, staying constant. For comparisons, we assume an idealized Dy- alloy family correspondingly. For the light rare-earth alloy series, $J$ and $\mu_{eff}$ are taken to be 4.5 and 3.52 $\mu_B$ respectively, corresponding to the Nd$^{3+}$ ion. For the heavy rare-earth alloy series, $J$ and $\mu_{eff}$ are taken to be 7.5 and 10 $\mu_B$ respectively, corresponding to the Dy$^{3+}$ ion. $T_D$ of the light rare-earth alloy series is assumed to be 352 K, which corresponds to Laves phase LaAl$_2$ [75]. For the heavy rare-earth alloy series, $T_D$ is assumed to be 384 K, which corresponds to Laves phase LuAl$_2$ [75]. $\Delta S_T(T,H)$ and $\Delta T_{ad}(T,H)$ can be calculated by Equation (5).

**Figure 2** (a) shows the $\Delta S_T(T)$ of the Nd- alloy family, and the maximum $\Delta S_T$ of the Dy- alloy family in magnetic fields of 5 T. In the temperature range of 77 ~ 20 K, the maximum $\Delta S_T$ of the light rare-earth alloy series are large. Especially in the vicinity of the condensation point of hydrogen (20 K), "giant" values are observed, being almost two times larger than those of the heavy rare-earth alloys with a $T_C$ near room temperature. **Figure 2** (b) shows the $\Delta T_{ad}(T)$ of the Nd- alloy family, and the maximum $\Delta T_{ad}$ of the Dy- alloy family in magnetic fields of 5 T. Near 20 K, the light rare-earth alloy series shows a considerable maximum $\Delta T_{ad}$, which is larger than or comparable to that of the heavy rare-earth alloys with a $T_C$ near room temperature. Both $\Delta S_T$ and $\Delta T_{ad}$ show an increasing maximum value with a decreasing $T_C$ in the temperature range of 77 ~ 20 K. As predicted by the calculations, we can see that the light rare-earth series also achieve a large $\Delta S_T$ and $\Delta T_{ad}$ at low cryogenic temperatures, especially in the vicinity of the condensation point of hydrogen.

The calculations above point out a way of designing a fully light rare-earth based magnetocaloric materials for hydrogen liquefaction: tune the $T_C$ down to 20 K. Though there are no such idealized alloy families where only $T_C$ varies, the chemical and physical similarities of the light rare-earth elements, namely Ce, Pr, and Nd, makes it easy to tune the $T_C$ of their alloys by mixing different rare-earth elements with different *de Gennes* factors on the rare-earth sublattices [76–78], as $T_C$ of rare-earth based alloys usually scales with *de Gennes* factor following the equation

$$T_C = \frac{2ZJG}{3k_B}, \quad (6)$$

where Z is the nearest neighbors, $J$ is the Heisenberg exchange constant, and G is the *de Gennes* factor [47].

It has been reported that Laves phases NdAl$_2$ with a $T_C$ near 77 K and PrAl$_2$ with a $T_C$ near 30 K are two magnetocaloric materials with a large $\Delta S_T$ and $\Delta T_{ad}$ [70]. However, these two compounds are unable to cover the full temperature range of 77 ~ 20 K. As shown in **Figure 1** (a), the *de Gennes* factor of Nd, Pr, and Ce are significantly different, decreasing from 1.84 for Nd to 0.18 for Ce. Based on the analyses above, we predict that a light rare-earth RAl$_2$ (R: rare-earth elements) Laves phase series that covers the full temperature range (77 ~ 20 K) required by magnetocaloric

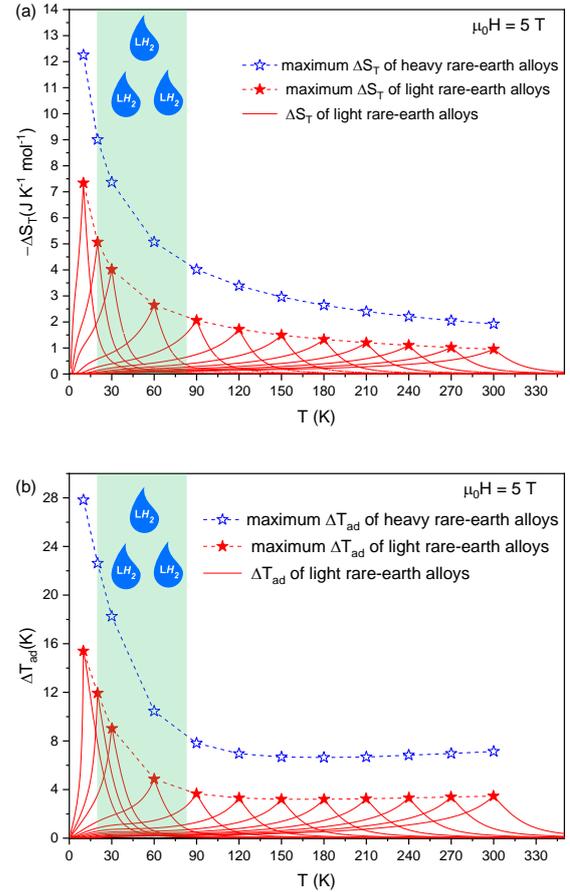

**Figure 2** (a) Magnetic entropy and (b) adiabatic temperature changes of a light rare-earth alloy series (idealized Nd- alloy family) and the maximum magnetic entropy and adiabatic temperature changes of the heavy rare-earth alloy series (idealized Dy- alloy family) from mean-field approach. Light green shadows marks the temperature range required by magnetocaloric hydrogen liquefaction.

hydrogen liquefaction can be realized by tuning the $T_C$ via mixing Pr and Nd in (Nd,Pr)Al$_2$, and Pr and Ce in (Pr,Ce)Al$_2$.

## 3. Experiment

Polycrystalline Nd$_x$Pr$_{1-x}$Al$_2$ (x = 1, 0.75, 0.5, 0.25) and Pr$_x$Ce$_{1-x}$Al$_2$ (x = 1, 0.75, 0.5) samples were synthesized by arc melting high-purity elements Ce (99.5 at. %), Pr (99.5 at%), Nd (99.5 at. %), and Al (99.998 at. %) under Ar atmosphere. We did not synthesize Pr$_{0.25}$Ce$_{0.75}$Al$_2$ and CeAl$_2$ since the later magnetization measurements show that the transition temperature of Pr$_{0.5}$Ce$_{0.5}$Al$_2$ is already below 20 K. To ensure good homogeneity, all the samples were melted three times. The ingots were turned upside down before each melting step. Evaporation of the rare-earth elements was negligible. Powder X-ray diffraction (XRD) patterns were collected at room temperature with an x-ray diffractometer (Stadi P, Stoe & Cie GmbH) equipped with a Ge[111]-Monochromator using Mo-K$_\alpha$-radiation in the Debye-Scherrer geometry. The XRD data were evaluated by Rietveld refinement with the FullProf software packages [79]. Backscatter electron (BSE) images were collected with a Tescan Vega 3 scanning electron microscope (SEM). A Physical Properties Measurement System (PPMS) from Quantum Design was used to measure the magnetization of the samples in magnetic fields up to 5 T. Heat capacity in magnetic fields of 0, 2, and 5 T was measured in the same PPMS with the $2\tau$ approach.

## 4. Results and discussions

### 4.1 Phase characterization

The Laves phases NdAl$_2$ and PrAl$_2$ crystallize in the MgCu$_2$ cubic structure (space group: 227). **Figure 3** (a) shows the XRD patterns of the (R$_1$,R$_2$)Al$_2$ (R$_1$: Nd, Pr, R$_2$: Pr, Ce) samples. Detailed Rietveld refinements are included in the supplementary. The Rietveld refinements confirm that (Nd,Pr)Al$_2$ and (Pr,Ce)Al$_2$ does not change their crystal structures with the variation of Nd, Pr, or Ce content. Phase fraction analyses demonstrate the high quality of all the samples since the impurities are undetectable. The lattice constants of (Nd,Pr)Al$_2$ and (Pr,Ce)Al$_2$ samples are plotted in **Figure 3** (b). The lattice constants increase almost linearly with increasing Pr content in (Nd,Pr)Al$_2$ series and increasing Ce content in (Pr, Ce)Al$_2$ series, respectively. This is coherent with the fact that CeAl$_2$ has the largest and NdAl$_2$ has the smallest lattice constant [78]. The quality of all the samples is further confirmed by SEM imaging. **Figure 3** (c) shows an example of the microscopy using BSE contrast, proving phase purity of the Pr$_{0.75}$Ce$_{0.25}$Al$_2$ sample.

### 4.2 Paramagnetic Curie temperature

**Figure 4** (a) plots the magnetization measurements of the Laves phases (R$_1$,R$_2$)Al$_2$ (R$_1$: Nd, Pr, R$_2$: Pr, Ce) as a function of temperature in magnetic fields of 5 T to check the saturation magnetization. The samples were firstly cooled down to 7 K in magnetic fields of 5 T, and then heated up with their

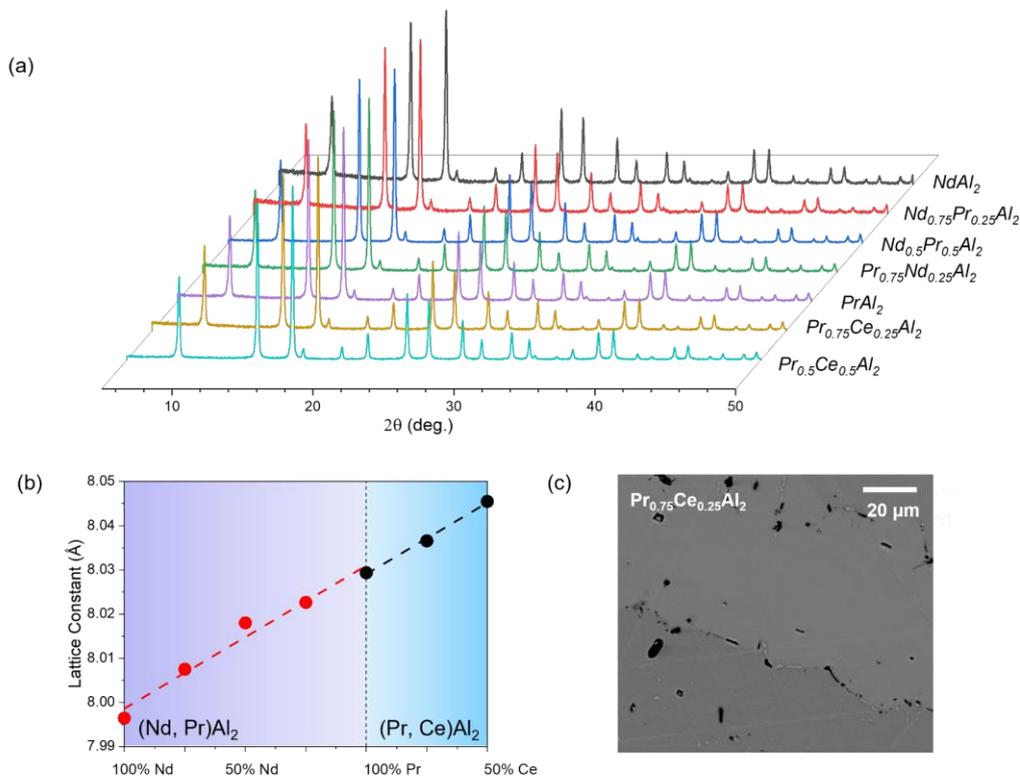

**Figure 3** (a) XRD patterns of the Laves phases (R$_1$,R$_2$)Al$_2$ (R$_1$: Nd, Pr, R$_2$: Pr, Ce) measured at room temperature. (b) Lattice constants of all the samples given by Rietveld refinement. The red dashed line is the linear fitting for (Nd, Pr)Al$_2$ samples, and the black line dashed line is the linear fitting for (Pr, Ce)Al$_2$ samples. (c) BSE image of Pr$_{0.75}$Ce$_{0.25}$Al$_2$. The black dots are holes.

magnetization measured in parallel. Under the same cooling and heating procedures, magnetization measurements in 0.1 T were done, the results are included in the supplementary. The saturated magnetization at 7 K of (Nd,Pr)Al$_2$ and Pr$_{0.75}$Ce$_{0.25}$Al$_2$ samples is rather close, in the range between 64 and 71 Am$^2$/kg. Pr$_{0.5}$Ce$_{0.5}$Al$_2$ is the only exception having a significantly lower magnetization of around 40 Am$^2$/kg.

**Figure 4** (b) shows the Curie-Weiss fits of all the samples in magnetic fields of 1 T. To reduce the deviation of the reciprocal magnetic susceptibility from the Curie-Weiss behavior, which may be associated with intrinsic factors such as the van Vleck effect, or extrinsic factors such as impurities [80–82], we performed the Curie-Weiss fit in 1 T. The paramagnetic Curie temperature can be determined by the intercepts of the Curie-Weiss fit with the x-axis [45]. The total effective magnetic moment $\mu_{eff}$ can be calculated by

$$\mu_{eff} = \frac{1}{\mu_B}\sqrt{\frac{3k_B M_R}{N_A \alpha}}, \qquad (7)$$

where $M_R$ is the molecular mass, $N_A$ the Avogadro constant, and $\alpha$ the slope of the linear fitting for $\mu_0 \chi^{-1}$ vs. $T$ [45].

**Figure 4** (c) plots the $\mu_{eff}$ of all the samples. The total effective magnetic moment $\mu_{eff}$ of (Nd,Pr)Al$_2$ increases roughly linearly with Pr content, whereas $\mu_{eff}$ of (Pr,Ce)Al$_2$ decreases with increasing Ce content. This observation can be explained by the fact that Pr$^{3+}$ has the largest magnetic moment of 3.58 $\mu_B$ and Ce$^{3+}$ has the smallest magnetic moment of 2.54 $\mu_B$ among the three light rare-earth ions.

**Figure 4** (d) plots the paramagnetic Curie temperature $\theta_p$ and the *de Gennes* factor $G$ of all the samples. The paramagnetic Curie temperature of (Nd,Pr)Al$_2$ decreases almost linearly with increasing Pr content, from 78.5 K for NdAl$_2$ to 32.6 K for PrAl$_2$, and it is the same case with (Pr,Ce)Al$_2$, from 32.6 K for PrAl$_2$ to 13.2 K for Pr$_{0.5}$Ce$_{0.5}$Al$_2$. This agrees with the decreasing trend of the *de Gennes* factor from NdAl$_2$ to Pr$_{0.5}$Ce$_{0.5}$Al$_2$ since as Equation (6) indicates, smaller *de Gennes* factor, lower Curie temperature. The paramagnetic Curie temperature is in good agreement with the values of the Curie temperatures of NdAl$_2$ and PrAl$_2$ given by Ref. [70]. The reported Curie temperature for NdAl$_2$ varies from 65 to 82 K, and that for PrAl$_2$ varies from 31 to 38.5 K [70, 78, 83–85].

In conclusion, by tuning *de Gennes* factors via mixing different rare-earth elements, a fully light rare-earth based magnetocaloric material system with a paramagnetic Curie temperature covering the temperature range required for magnetocaloric hydrogen liquefaction (77 ~ 20 K) is developed. The large effective magnetic moments $\mu_{eff}$ are retained from NdAl$_2$ to Pr$_{0.75}$Ce$_{0.25}$Al$_2$.

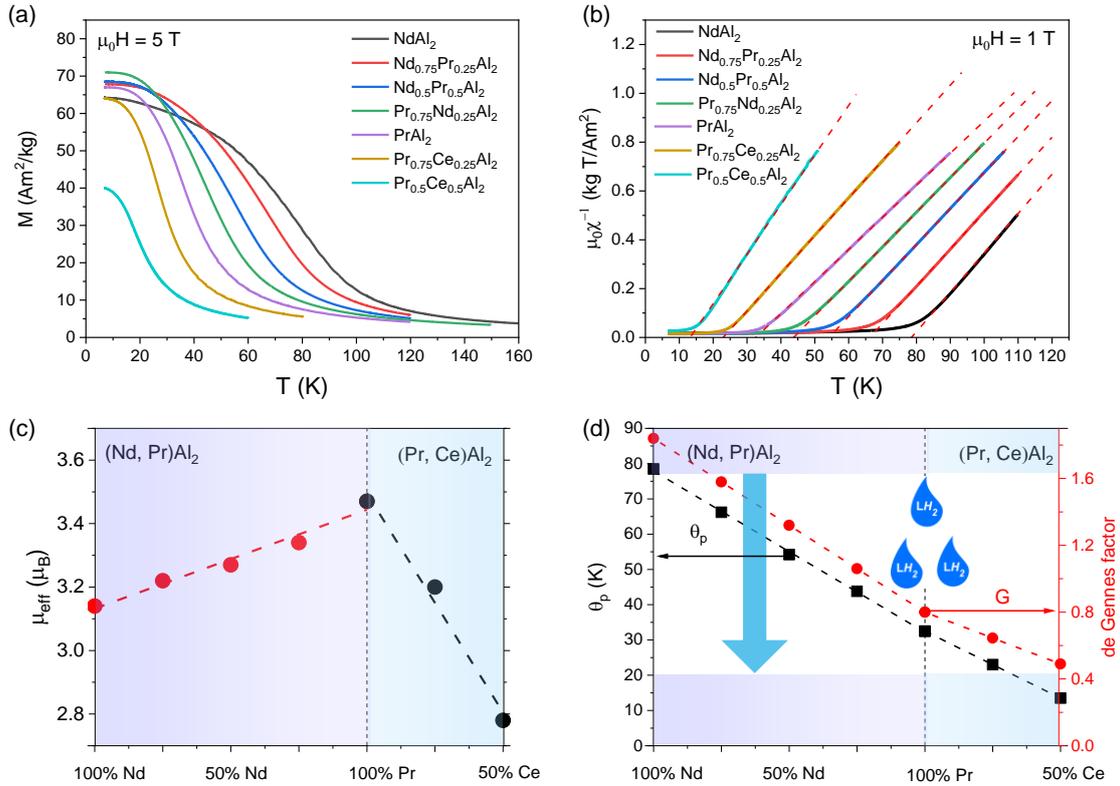

**Figure 4** (a) Magnetizations of the Laves phases (R$_1$,R$_2$)Al$_2$ (R$_1$: Nd, Pr, R$_2$: Pr, Ce) as a function of temperature in magnetic fields of 5 T. (b) Curie-Weiss fits of the Laves phases (R$_1$,R$_2$)Al$_2$ (R$_1$: Nd, Pr, R$_2$: Pr, Ce) in magnetic fields of 1 T. (c) Effective magnetic moments of the samples using the Curie-Weiss law. (d) Paramagnetic $T_C$ of the samples determined by the Curie-Weiss law (black squares) and the average *de Gennes* factor (red circles).

### 4.3 Magnetocaloric effect

According to the theoretical calculations above, the maximum $\Delta S_T$ of the light rare-earth based magnetocaloric materials are expected to be large in the vicinity of the condensation point of hydrogen (20 K). **Figure 5** (a) ~ (g) plot the $\Delta S_T$ of all the samples in magnetic fields of 2 and 5 T. $\Delta S_T(T, \Delta H)$ is determined by two methods: (1) heat capacity measurements under constant applied fields via the thermodynamic equation $S(T, \Delta H) - S(T, 0) = \int_0^T \frac{C_p(T,\Delta H) - C_p(T,0)}{T} dT$ ( $C_p$ is the isobaric heat capacity) [73, 86], (2) magnetization vs. fields measurements via the Maxwell relation $\Delta S_T(T, \Delta H) = \int_{H_0}^{H_1} \frac{\partial M(H,T)}{\partial T} dH$ [37, 87, 88]. An example of how the $\Delta S_T$ is calculated by these two methods for PrAl$_2$ is included in the supplementary. Both methods fit well, as the points of the $\Delta S_T$ from magnetization measurements mostly lie on the lines of the $\Delta S_T$ from heat capacity measurements. Besides, the temperatures where $\Delta S_T$ peaks are close to the paramagnetic Curie temperatures, consistent with the feature that second-order magnetocaloric materials show a maximum $\Delta S_T$ near their Curie temperatures [36, 65].

In agreement with the calculations in **Figure 2** (a) above, the maximum $\Delta S_T$ increases from 7.21 J kg$^{-1}$ K$^{-1}$ for NdAl$_2$ to 18.53 J kg$^{-1}$ K$^{-1}$ for Pr$_{0.75}$Ce$_{0.25}$Al$_2$ in magnetic fields of 5 T, and from 3.67 J kg$^{-1}$ K$^{-1}$ for NdAl$_2$ to 10.48 J kg$^{-1}$ K$^{-1}$ for Pr$_{0.75}$Ce$_{0.25}$Al$_2$ in magnetic fields of 2 T. We observe an exception in this material series that Pr$_{0.5}$Ce$_{0.5}$Al$_2$ has a smaller maximum $\Delta S_T$ than that of PrAl$_2$ and Pr$_{0.75}$Ce$_{0.25}$Al$_2$. Similar observations were reported in RNi$_2$ (R: Gd, Tb, Dy, Ho, Er) [45, 71] and RAl$_2$ (R: Gd, Tb, Dy, Ho, Er, Tm) [72] series. One contribution to this decrease is the reduction of the effective magnetic moment $\mu_{eff}$. As revealed in **Figure 4** (c), Pr$_{0.5}$Ce$_{0.5}$Al$_2$ has the lowest $\mu_{eff}$ in the series. Another contribution might be due to the fact that crystalline electric field has a considerable influence on the magnetocaloric effect in low temperatures [45]. In the case of Pr$_{0.5}$Ce$_{0.5}$Al$_2$, the crystalline electric field may decrease the magnetic entropy change. However, this speculation needs to be validated.

**Figure 5** (h) and (i) compares the maximum $\Delta S_T$ of the light rare-earth RAl$_2$ series in this work to some of the other light rare-earth magnetocaloric materials [89–94] and the heavy rare-earth RAl$_2$ (R: Tb, Dy, Ho, Er) and Tb$_x$Ho$_{1-x}$Ni$_2$ (x = 0.25, 0.5, 0.75, 1) series [45, 95, 96]. Form the plots, we see

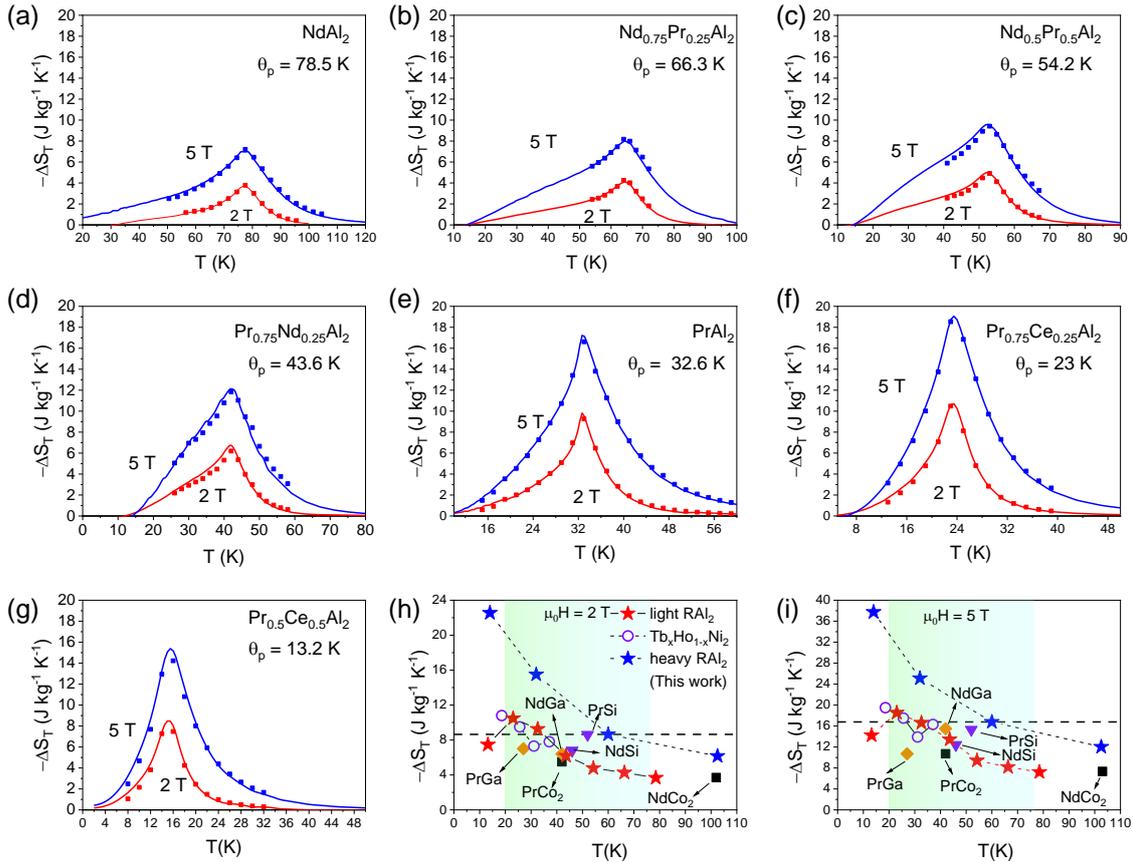

**Figure 5** (a) ~ (g) -$\Delta S_T$ of (Nd, Pr)Al$_2$ and (Pr, Ce)Al$_2$ in magnetic fields of 2 T and 5 T. The solid lines are the -$\Delta S_T$ calculated from heat capacity measurements, and the squares are the ones calculated from magnetization vs. magnetic field measurement. (h) ~ (i) Comparisons of the maximum -$\Delta S_T$ of the light rare-earth series in this work, RCo$_2$ (Ref. [89, 90]), RSi (Ref. [91, 92]), and RGa (Ref. [93, 94]) where R = Pr, Nd, the heavy rare-earth based RAl$_2$ (Ref. [45, 95]) where R = Tb, Dy, Ho and Er, and Tb$_x$Ho$_{1-x}$Ni$_2$ (x = 0.75, 0.25, 0.5, 0)(Ref. [61]) in magnetic fields of 2 and 5 T. The green shadows mark the temperature range required for magnetocaloric hydrogen liquefaction (77 K ~ 20 K) and the black dashed lines highlight the values of DyAl$_2$.

that the light rare-earth RAl$_2$ series is highly competitive compared to the other light rare-earth magnetocaloric materials. Pr$_{0.75}$Ce$_{0.25}$Al$_2$ exhibits the largest maximum $\Delta S_T$ among all the light rare-earth magnetocaloric materials in **Figure 5** (h) and (i). The rest light rare-earth RAl$_2$ samples show a maximum $\Delta S_T$ that is larger than or comparable to the other light rare-earth compounds with a similar ordering temperature (except PrSi which shows a $\Delta S_T$ larger than PrAl$_2$ and Pr$_{0.75}$Nd$_{0.25}$Al$_2$ samples).

Although the heavy rare-earth series RAl$_2$ (R: Tb, Dy, Ho, Er) shows an obvious larger maximum $\Delta S_T$ than their light rare-earth counterparts in the vicinity of their ordering temperatures, the light rare-earth RAl$_2$ series exhibits large $\Delta S_T$ near hydrogen condensation point (20 K), with Pr$_{0.75}$Ce$_{0.25}$Al$_2$ having a larger $\Delta S_T$ than DyAl$_2$, and PrAl$_2$ having a significantly larger $\Delta S_T$ than TbAl$_2$. Compared to other heavy rare-earth Laves phases such as Tb$_x$Ho$_x$Ni$_2$ (x = 0.25, 0.5, 0.75, 1) [45, 95, 96], Pr$_{0.75}$Ce$_{0.25}$Al$_2$, PrAl$_2$ and Pr$_{0.75}$Nd$_{0.25}$Al$_2$ show a maximum $\Delta S_T$ that is larger than or comparative to that of Tb$_x$Ho$_{1-x}$Ni$_2$.

Adiabatic temperature change $\Delta T_{ad}$ is as important as $\Delta S_T$ for magnetocaloric effect. **Figure 6** (a)~(g) plot the $\Delta T_{ad}$ of the light rare-earth RAl$_2$ samples determined by equation (5) via constructing the $S(H,T)$ curves from the heat capacity measurements in magnetic fields of 0, 2, and 5 T. In agreement with the theoretical calculations above, the maximum $\Delta T_{ad}$ of the light rare-earth RAl$_2$ series is large in the vicinity of the condensation point of hydrogen (20 K), with all the three (Pr,Ce)Al$_2$ samples showing a maximum $\Delta T_{ad}$ over 2 K in magnetic fields of 2 T. In 5 T, all the three (Pr,Ce)Al$_2$ samples have a maximum adiabatic temperature change over 4 K, and Pr$_{0.75}$Ce$_{0.25}$Al$_2$ even shows a value close to 5 K. **Figure 6** (h)~(i) compares the maximum $\Delta T_{ad}$ of the light- and heavy rare-earth RAl$_2$, and the Laves phases Tb$_x$Ho$_{1-x}$Ni$_2$. The light rare-earth RAl$_2$ shows a maximum $\Delta T_{ad}$ which is about 1/3 ~ 1/2 of the maximum values of the heavy rare-earth RAl$_2$ in the vicinity of their ordering temperatures. Compared to Tb$_x$Ho$_{1-x}$Ni$_2$ [45, 95, 96] in magnetic fields of 2 T, PrAl$_2$ and Pr$_{0.75}$Nd$_{0.25}$Al$_2$ show a maximum $\Delta T_{ad}$ that is about half of the values of HoNi$_2$ and Tb$_{0.25}$Ho$_{0.75}$Ni$_2$, but comparative to that of Tb$_{0.5}$Ho$_{0.5}$Ni$_2$ and Tb$_{0.75}$Ho$_{0.25}$Ni$_2$.

In summary, the light rare-earth RAl$_2$ alloy series shows a large $\Delta S_T$ being comparable to the other light rare-earth based materials in **Figure 5** (h) and (i). Both of their $\Delta S_T$ and $\Delta T_{ad}$ show a large maximum value near 20 K. Despite showing a smaller $\Delta S_T$ and $\Delta T_{ad}$ than their heavy rare-earth

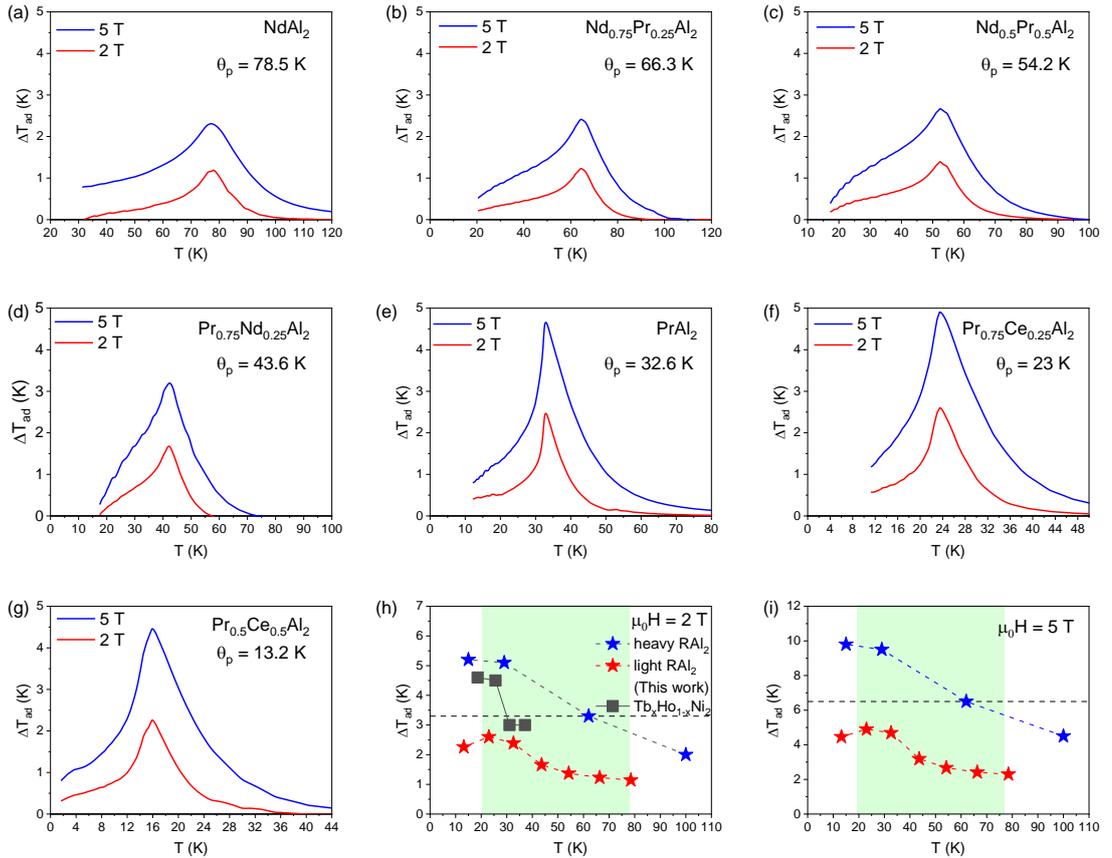

**Figure 6** (a)~(g) $\Delta T_{ad}$ of (R$_1$,R$_2$)Al$_2$ (R$_1$: Nd, Pr, R$_2$: Pr, Ce) determined from heat capacity measurements in magnetic fields of 2 and 5 T. (h)~(i) comparisons of the $\Delta T_{ad}$ of the light- and heavy rare-earth RAl$_2$ series, and the Tb$_x$Ho$_{1-x}$Ni$_2$ (x = 0.75, 0.25, 0.5, 0) (data of heavy rare-earth RAl$_2$ (R = Tb, Dy, Ho, and Er) and Tb$_x$Ho$_{1-x}$Ni$_2$ are taken from Ref. 45 and 61, respectively). The green shadows mark the temperature range required for magnetocaloric hydrogen liquefaction (77 K ~ 20 K) and the black dashed lines highlight the values of DyAl$_2$.

counterparts in the vicinity of their ordering temperatures, we see a high potential of the light rare-earth RAl$_2$ series for magnetocaloric hydrogen liquefaction, especially near 20 K, the condensation point of hydrogen.

## 5. Conclusions

The relatively high abundance of light rare-earth elements in the earth's crust makes their alloys appealing for magnetocaloric hydrogen liquefaction on an industrial scale. In this work, we aimed at designing a fully light rare-earth based magnetocaloric material system covering the full temperature range from 77 to 20 K required by hydrogen liquefaction.

In order to formulate a strategy for alloy design, the mean-field approach shown in our previous work is further developed to be applied to the light rare-earth alloys. From the theoretical analysis, we see that if $T_C$ of a light rare-earth based magnetocaloric material is tuned towards lower cryogenic temperature, the magnetocaloric effect is supposed to become stronger. Especially in the vicinity of the condensation point of hydrogen, the mean-field approach predicts significantly large $\Delta S_T$ and $\Delta T_{ad}$. Based on these observations and taking the chemical and physical similarities of the light rare-earth elements, a design strategy for developing light rare-earth material series for hydrogen liquefaction is used: tune the $T_C$ by mixing the light rare-earth elements with different de Gennes factors.

Consequently, a light rare-earth RAl$_2$ Laves phase series with the ordering temperatures covering the temperature range from 77 to 20 K is successfully developed. This material series exhibits large $\Delta S_T$. Especially near 20 K (condensation point of hydrogen), the (Pr$_x$, Ce$_{1-x}$)Al$_2$ (x = 0, 0.75, 0.5) samples show a $\Delta S_T$ that is larger than or comparable to that of DyAl$_2$, a heavy rare-earth based magnetocaloric material which is often proposed to be used in an active regenerator for hydrogen liquefaction [32]. Large $\Delta T_{ad}$ in the vicinity of 20 K are achieved in the (Pr$_{1-x}$Ce$_x$)Al$_2$ (x = 0, 0.75, 0.5) samples, which show a value that is more than two third of that of DyAl$_2$.

This design strategy for designing the light rare-earth RAl$_2$ Laves phase series for magnetocaloric hydrogen liquefaction may be applied to other light rare-earth alloys to tailor their magnetocaloric effects for the liquefaction of industrial gases, inclusive but not limited to hydrogen gas. In addition, our work is also helpful for designing magnetocaloric composites, since tuning the Curie temperature in layered structures with a constant $\Delta S_T$ over a wide temperature range is important for applications [60, 62].

## 6. Acknowledgement

We appreciate the financial supports from Helmholtz Association via the Helmholtz-RSF Joint Research Group (Project No. HRSF-0045), from the HLD at HZDR (member of the European Magnetic Field Laboratory (EMFL)). We further gratefully acknowledge supports from the DFG through the Würzburg-Dresden Cluster of Excellence on Complexity and Topology in Quantum Matter-*ct.qmat* (EXC 2147, Project ID 39085490), the CRC/TRR 270 (Project-ID 405553726) and the Project-ID 456263705, from the ERC under the European Union's Horizon 2020 research and innovation program (Grant No. 743116, Cool Innov), and the Clean Hydrogen Partnership and its members within the framework of the project HyLICAL (Grant No. 101101461).

## 7. Data availability statements

The data that support the findings of this study are available upon reasonable request from the authors.

Supplementary

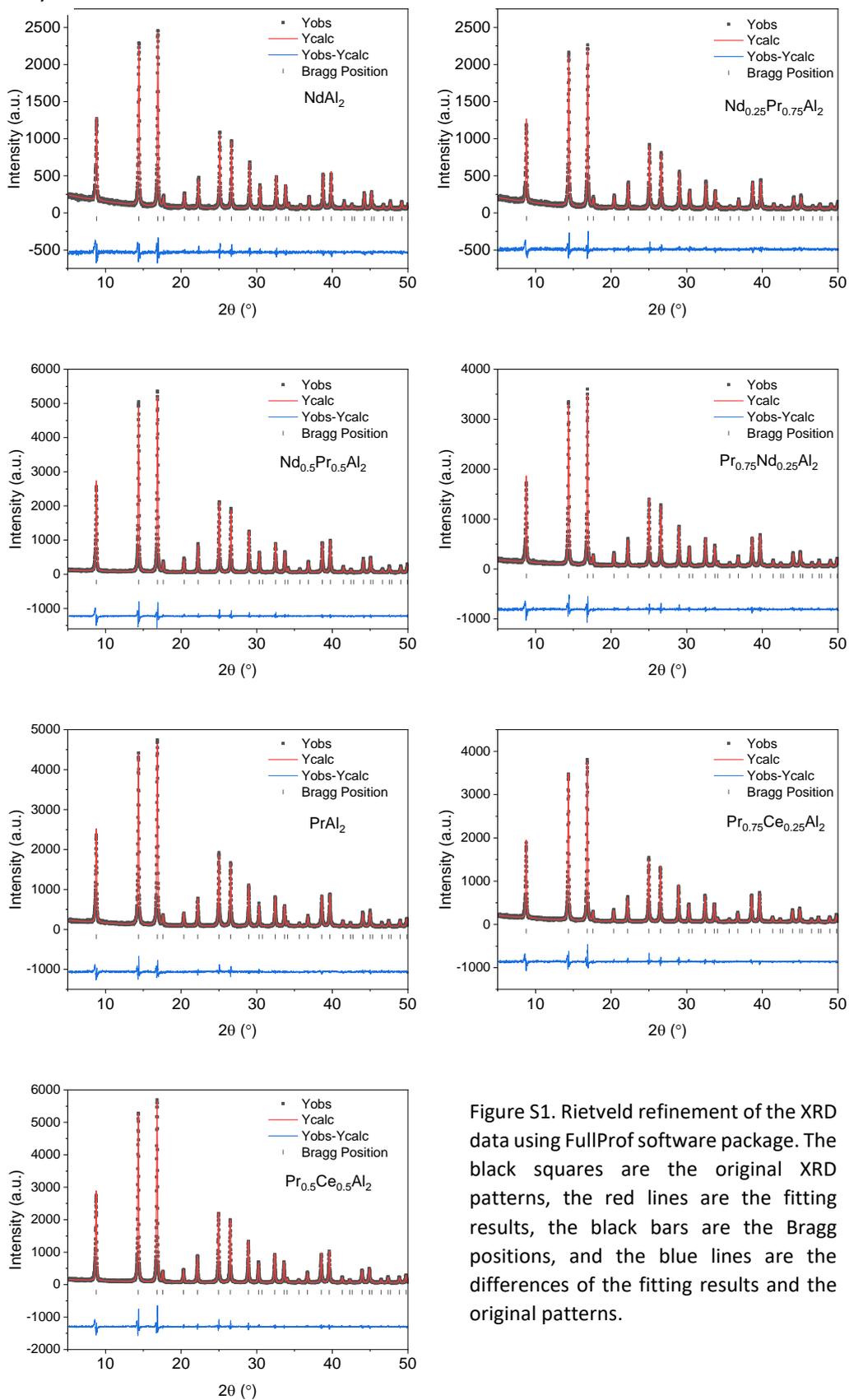

Figure S1. Rietveld refinement of the XRD data using FullProf software package. The black squares are the original XRD patterns, the red lines are the fitting results, the black bars are the Bragg positions, and the blue lines are the differences of the fitting results and the original patterns.

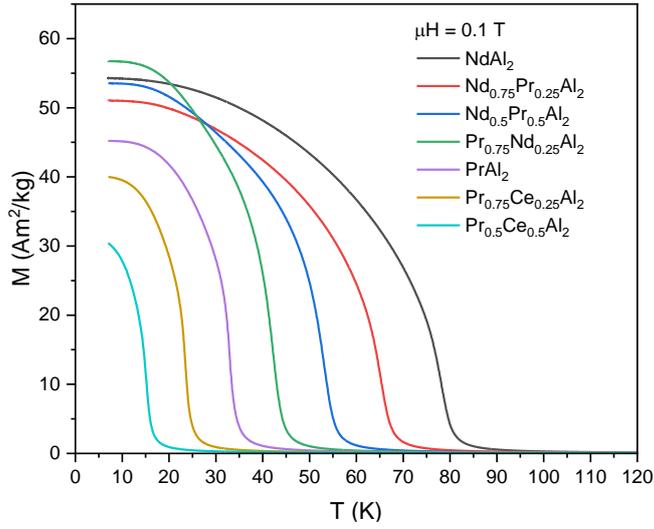

Figure S2. Magnetization as a function of temperatures for all the light rare-earth RAl$_2$ samples in magnetic fields of 0.1 T.

As shown in Figure S3, the total entropy $S(T)$ curves of PrAl$_2$ in magnetic fields of 0, 2, and 5 T were constructed by the heat capacity measurements via the equation $S(T) = \int_0^T \frac{C_p}{T} dT$. The $\Delta S_T$ and $\Delta T_{ad}$ from the heat capacity measurements were calculated by the equation (5) shown in the main article.

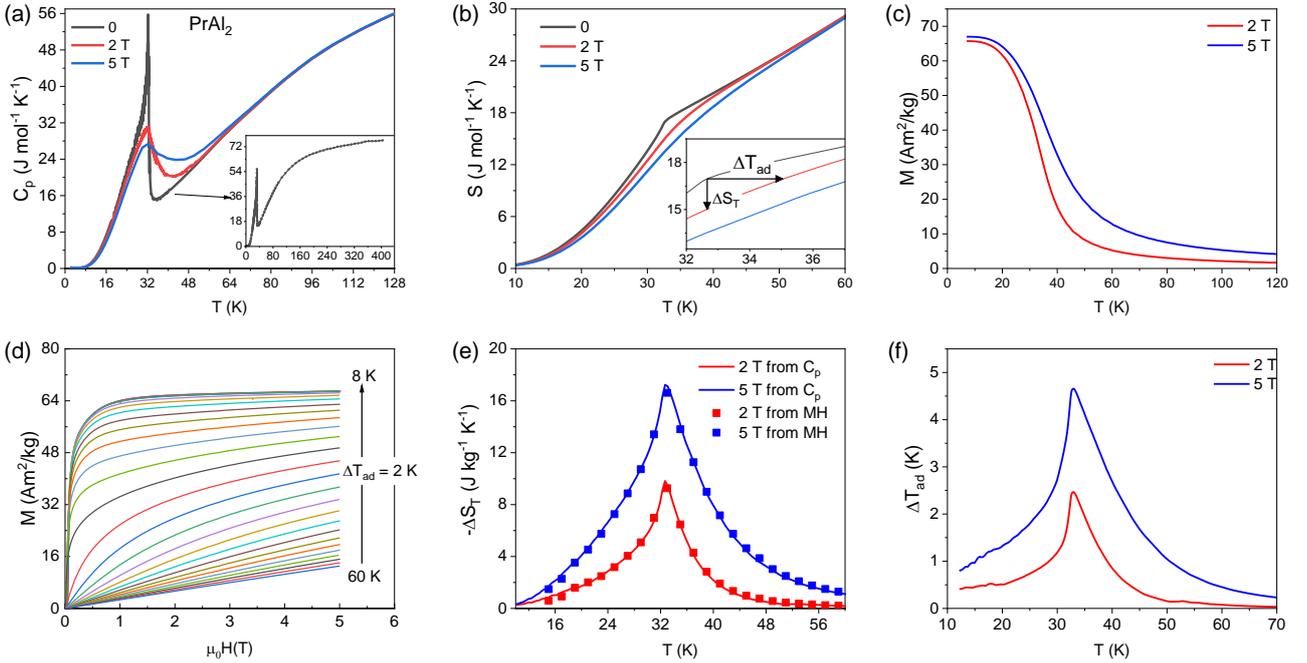

Figure S3. (a) heat capacity of PrAl$_2$ measured in 0, 2, and 5 T. The inset shows the heat capacity in zero fields up to 404 K. (b) $S(T)$ curves of PrAl$_2$ in magnetic fields of 0, 2, and 5 T. The inset shows the curves in temperature range of 32 ~ 37 K. (c) Magnetization of PrAl$_2$ as a function of temperature in magnetic fields of 2 and 5 T. (d) Magnetization of PrAl$_2$ as a function of magnetic fields at a constant temperature from 8 K to 60 K with a step of 2 K. (e) Isothermal magnetic entropy change of PrAl$_2$. The solid lines are calculated from heat capacity measurements while the squares are calculated from magnetization vs. magnetic field (MH) at a constant temperature. (f) adiabatic temperature change of PrAl$_2$ calculated from heat capacity measurements.